\documentclass[onecollarge,natbib]{svjour2}
\bibpunct{[}{]}{;}{n}{}{,} 
\smartqed  
\usepackage{graphicx}
\usepackage[normalem]{ulem}
\usepackage{amsmath}
\usepackage{amssymb}
\usepackage{mathtools}
\usepackage{color}
\usepackage{xcolor}
\usepackage{lineno}
\usepackage{hyperref}
\usepackage{relsize}
\usepackage{mathrsfs}
\usepackage{bm}
\usepackage{cancel}
\usepackage{soul}
\usepackage{booktabs}
\usepackage{caption}
\usepackage[caption=false]{sub fig} 
\colorlet{darkgreen}{green!50!black}
\colorlet{brightyellow}{yellow!75!red}
\colorlet{orange}{red!50!yellow}
\colorlet{darkblue}{blue!60!black}
\colorlet{darkred}{red!80!black}

\journalname{Few-Body Systems}
\begin{document}
 
\title{Trends and Progress in Nuclear and Hadron Physics: a straight or winding road
       \thanks{Presented at LightCone 2016, Lisbon, Portugal.}
}

\titlerunning{BLFQ - Progress and Prospects}        

\author{James~P.~Vary   \and
        Lekha Adhikari        \and
	Guangyao~Chen    \and
	Meijian~Li	              \and 
        Yang~Li                   \and
        Pieter~Maris           \and
        Wenyang~Qian      \and
        John~R.~Spence   \and
        Shuo Tang              \and
        Kirill~Tuchin	     \and
        Xingbo~Zhao
}

\authorrunning{Vary et al.} 

\institute{James~P.~Vary \and Lekha Adhikari \and Guangyao~Chen
\and Meijian~Li \and Yang~Li  \and  Pieter~Maris 
\and Wenyang~Qian \and John~R.~Spence  
\and Shuo Tang \and Kirill~Tuchin \and Xingbo~Zhao 
\at Department of Physics and Astronomy, Iowa State University, Ames, IA 50011, USA \\
\email{jvary@iastate.edu} 
\and
           Xingbo~Zhao \at
           Institute of Modern Physics, Chinese Academy of Sciences, Lanzhou, 730000, China \\
}

\date{\today}

\maketitle

\begin{abstract}

Quantitative calculations of the properties of hadrons and nuclei, with assessed uncertainties, have emerged as competitive with experimental measurements in a number of major cases. We may well be entering an era where theoretical predictions are critical for experimental progress. Cross-fertilization between the fields of relativistic hadronic structure and non-relativistic nuclear structure is readily apparent.  Non-perturbative renormalization methods such as Similarity Renormalization Group and Okubo-Lee-Suzuki schemes as well as many-body methods such as Coupled Cluster, Configuration Interaction and Lattice Simulation methods are now employed and advancing in both major areas of physics.  New algorithms to apply these approaches on supercomputers are shared among these areas of physics.  The roads to success have intertwined with each community taking the lead at various times in the recent past.  I briefly sketch these fascinating paths and comment on some symbiotic relationships. I also overview some recent results from the Hamiltonian Basis Light-Front Quantization approach.  
\keywords{Non-Perturbative Physics \and Computational Physics \and Basis Function Method}
\end{abstract}


\section{Introduction}\label{introduction}

Nuclear and hadronic physics share methods for non-perturbatively solving the quantum many-body
problem. Lattice, wave-equation and Hamiltonian eigenvalue methods represent major many-body 
approaches where developments in one subfield have impacts on the other. Renormalization and 
effective field theory techniques are also under development in both subfields. For these reasons,
researchers often enjoy ``dual citizenship'' in these subfields. Advances in computational
physics and high-performance computing have spurred great progress in nuclear and 
hadronic physics by enabling solutions with unprecedented accuracy and quantified uncertainties.

In nuclear physics, the development of effective interactions has a long history. 
The pioneering works of Brueckner, Bethe and Goldstone (see Ref. \cite{BRANDOW:1967zz}
for a review) led to decades of successful applications by combining non-perturbative and perturbative approaches.
Eventually, more robust non-perturbative methods such as Okubo-Lee-Suzuki (OLS) (see Ref. \cite{Barrett:2013nh} for a 
review) and the Unitary Correlation Operator Method (UCOM) (see Ref. \cite{Roth:2010bm} for a 
review) paved another road to extensive successes.

These methods were developed to enable many-body Hamiltonian approaches to achieve improved
convergence in nuclear physics within available computational methods and resources.  The methods
improved and the resources grew simultaneously so that increasingly precise solutions have
been attained both for a growing number of nuclei and for a wider array of experimental observables \cite{Barrett:2013nh}.

In the hardron arena, we have witnessed parallel rapid developments of the Hamiltonian approach 
and some developments have been intertwined with those in nuclear physics.  Based upon Dirac's front form of relativistic Hamiltonian dynamics \cite{Dirac:1949cp}, hadronic physics witnessed the development and application of the practical Discretized Light Cone Quantization 
(DLCQ) approach (see Ref. \cite{Brodsky98.299} for a review). On the renormalization frontier, we saw the 
introduction and application of the Similarity Renormalization Group (SRG) technique \cite{Glazek:1993rc,Wegner1994} which has enjoyed remarkable success in nuclear physics as well (see Ref. \cite{Bogner:2007rx} and references therein).

A major example of a many-body method that has enjoyed success in an array of scientific fields is the Coupled-Cluster method of Coester and Kuemmel \cite{Coester1958,Coester1960}.  From its genesis within nuclear physics, it has been actively developed and applied in quantum chemistry (see Ref. \cite{Kuemmel2003} for a review), nuclear physics (see Ref. \cite{Hagen:2013nca} for a review) and it has also been introduced into light-front field theory with promising results \cite{Hiller:2016itl}.

\section{Basis Function Approach} \label{sec 2}

Given these rapid advances in many-body methods, renormalization approaches, algorithms for
simulations and high-performance computers sketched above, it is appropriate to survey 
a few applications that have emerged from these advances. Due to space limitations, 
I will select a few highlights of recent results from development and applications of 
Basis Light-Front Quantization (BLFQ) 
\cite{Vary:2009gt,Honkanen:2010rc,Vary:2011np}
and its time-dependent generalization tBLFQ
\cite{Zhao:2013jia,Zhao:2013cma}. For a more complete overview of 
current and planned research in light-front Hamiltonian theory, 
see Ref.~\cite{Bakker2013.165} and references therein.

Simply stated, our goal is to solve the Hamiltonian eigenvalue problem
expressed as, 
  \begin{linenomath*}
  \begin{equation}
   \hat P^+ \hat P^- |\psi_h\rangle = M^2_h  |\psi_h\rangle
  \end{equation}
  \end{linenomath*}
where $\hat P^\pm = \hat P^0 \pm \hat P^3$ are the longitudinal momentum ($+$) and the light-front quantized Hamiltonian operator ($-$).
The invariant-mass spectrum results from the product of their eigenvalues. The eigenstates $|\psi_h\rangle$, or light-front wavefunctions (LFWFs) in either coordinate or momentum space, yield predictions for hadron structures in deep inelastic scattering (DIS) \cite{Brodsky98.299}, deeply virtual Compton scattering (DVCS) \cite{Brodsky:2000xy} and diffractive hadron production in ultra-peripheral heavy ion collisions \cite{Afanasiev:2009hy,Abbas:2013oua}, among many other applications.

The Fock space expansion for $|\psi_h\rangle$ produces a sparse-matrix many-body eigenvalue problem.
BLFQ was introduced to implement azimuthal rotational symmetry and to accelerate convergence in bound state applications, especially with confining interactions as in QCD. By choosing the two-dimensional (2D) 
harmonic-oscillator (HO) for the transverse modes, one benefits from the developments of the no-core shell model (NCSM) used successfully in nuclear many-body theory \cite{Barrett:2013nh,Navratil:2000ww,Navratil:2000gs,Maris:2008ax} while retaining a fully covariant framework. The ability to factorize
transverse center-of-mass motion in the LFWFs in order to preserve transverse boost invariance is one of the appealing features of BLFQ~\cite{Vary:2009gt,Li:2013cga,Maris:2013qma}. For applications to bound state problems in QCD, we note that the choice of the 2D-HO for the transverse basis space is harmonious with the phenomenologically successful light-front AdS/QCD soft-wall Hamiltonian for the hadrons \cite{Brodsky09.081601,Brodsky15.1} .

BLFQ has been used to solve QED problems at strong coupling such as the electron anomalous 
magnetic moment  
\cite{Honkanen:2010rc,Zhao:2014xaa}, non-linear Compton scattering 
\cite{Zhao:2013jia,Zhao:2013cma} 
and the positronium spectrum \cite{Wiecki:2014ola}.
In this paper, we briefly summarize recent BLFQ applications to bound-state and scattering problems:
Electron form factors~\cite{Tang:2017aaa} in Sect.~\ref{sec 3},
Yukawa model~\cite{Qian:2017aaa} in Sect.~\ref{sec 4},
heavy quarkonium \cite{Li:2015zda} in Sect.~\ref{sec 5}, 
vector meson production in Sect.~\ref{sec 6}, and
the extension to tBLFQ~\cite{Zhao:2013jia,Zhao:2013cma,Chen:2016aaa} in Sect.~\ref{sec 7}, 
each of which points pathways to future developments and applications.
A promising approach to non-perturbative renormalization is also
presented at this meeting~\cite{Zhao:2014hpa,Zhao_this_meeting}. 

%
%
\section{Electron form factors in BLFQ} \label{sec 3}

\begin{figure}
\centering 
\includegraphics[width=1.0\textwidth]{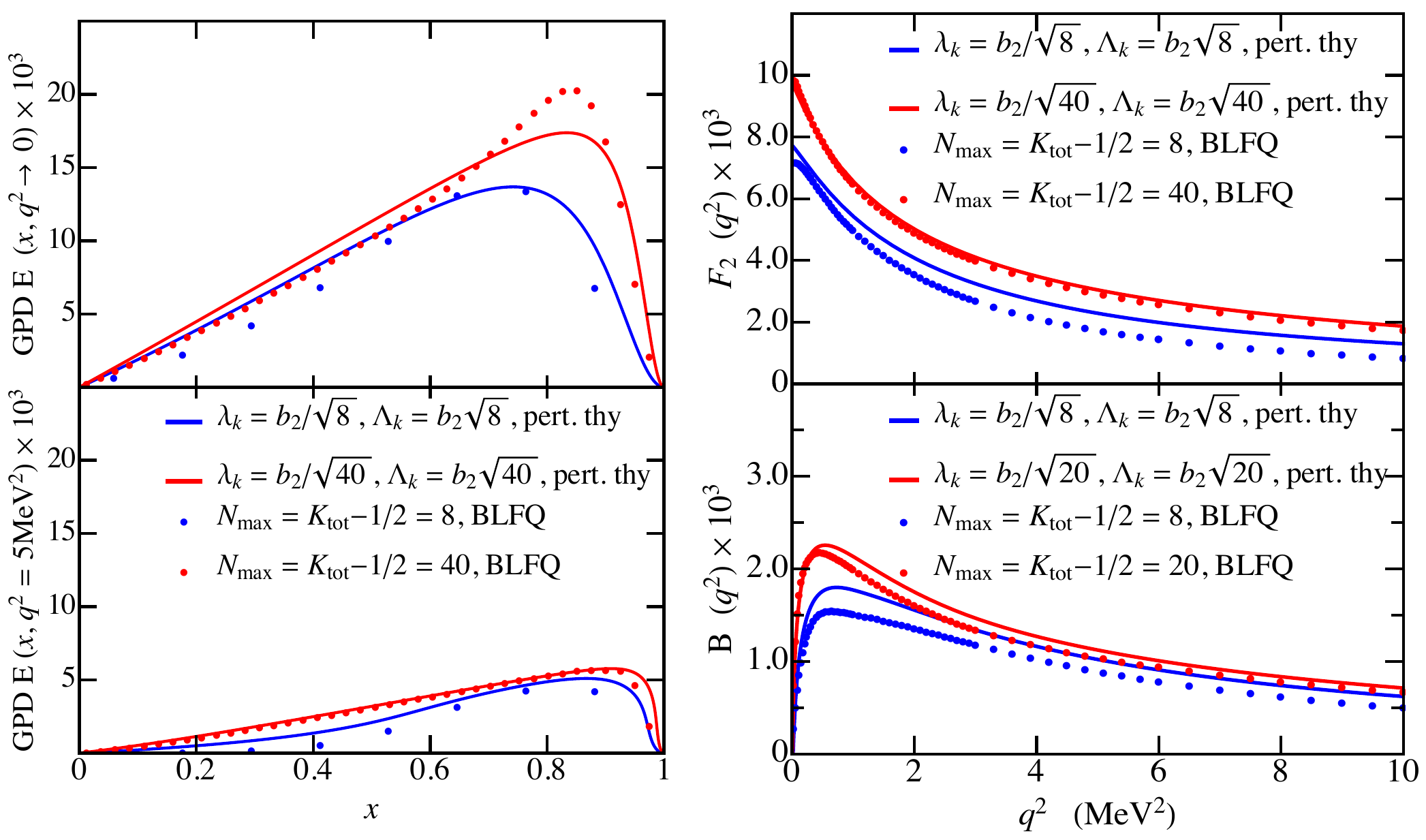}
\caption{(Color online) 
Electron form factors obtained from LFPT (solid lines) and BLFQ (dotted curves) 
at infrared (ultraviolet) regulator $\lambda_k$ ($\Lambda_k$) matched to
BLFQ regulators $N_{max}$ and $K_{tot}$.
\textit{Left}: Comparison of GPD E$(x,q^2)$, for $q^2=0$ (top panel) 
and $q^2=5 $ MeV$^2$ (bottom panel). 
The integral of GPD E$(x,q^2)$ over the momentum fraction x contributes to $F_2$. 
\textit{Right}: Form factors as functions of $q^2$. 
Top panel: Pauli form factor $F_2(q^2)$, bottom panel: gravitomagnetic form factor 
$B(q^2) = B_f + B_b$.
}
\label{electronff}
\end{figure}

Following earlier applications to QED where we evaluated the electron
anomalous magnetic moment in BLFQ and compared the results with
perturbation theory \cite{Honkanen:2010rc,Zhao:2014xaa}, we now
evaluate the electron's Generalized Parton Distribution (GPD) E$(x,q^2)$,
Pauli form factor $F_2(q^2)$, and anomalous gravitomagnetic form factor 
$B(q^2) = B_f + B_b$.
These are displayed in Fig. \ref{electronff} with values for the 
regulators provided in the legends. Here, $\lambda_k$ ($\Lambda_k$)
signifies the infrared (ultraviolet) cutoff matched to the BLFQ regulators
which are defined in terms of $b_2$ and $N_{\rm max}$. The quantity  
$ b_2=b/ \sqrt{2}$ represents the 2D-HO parameter of relative motion 
and $b=M$ is the 2D-HO parameter in the one-particle sector.
The regulator $N_{\rm max}$ defines the upper limit of the sum of 2D-HO quanta
for the transverse basis states and the regulator $K=K_{tot}$ defines the upper limit of the 
longitudinal plane wave modes for where we adopt antiperiodic (periodic) boundary 
conditions for fermions (bosons); i.e. $K=K_{tot}$ is the sum of the electron's
 (half-odd-integer) and the photon's (integer) longitudinal quanta .
In addition to showing the expected agreement between light-front perturbation theory (LFPT)
and DLCQ, we re-confirm the vanishing of the electron anomalous gravitomoment, i.e. 
$B(q^2=0) = 0$ \cite{Brodsky:2000ii}.

%

%
%
\section{Yukawa model in BLFQ} \label{sec 4}

\begin{figure}
\centering 
\includegraphics[width=.49\textwidth]{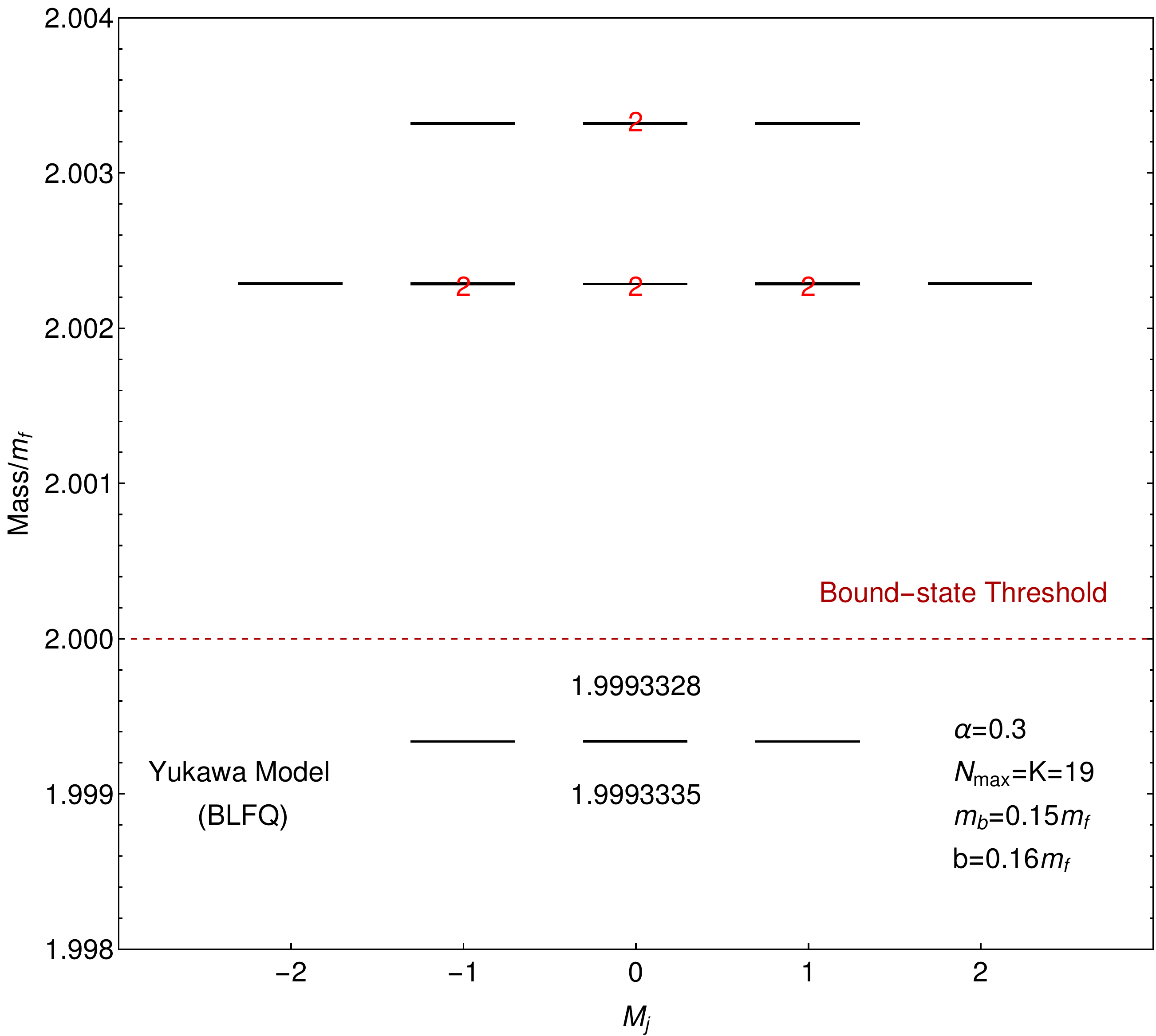}
\includegraphics[width=.49\textwidth]{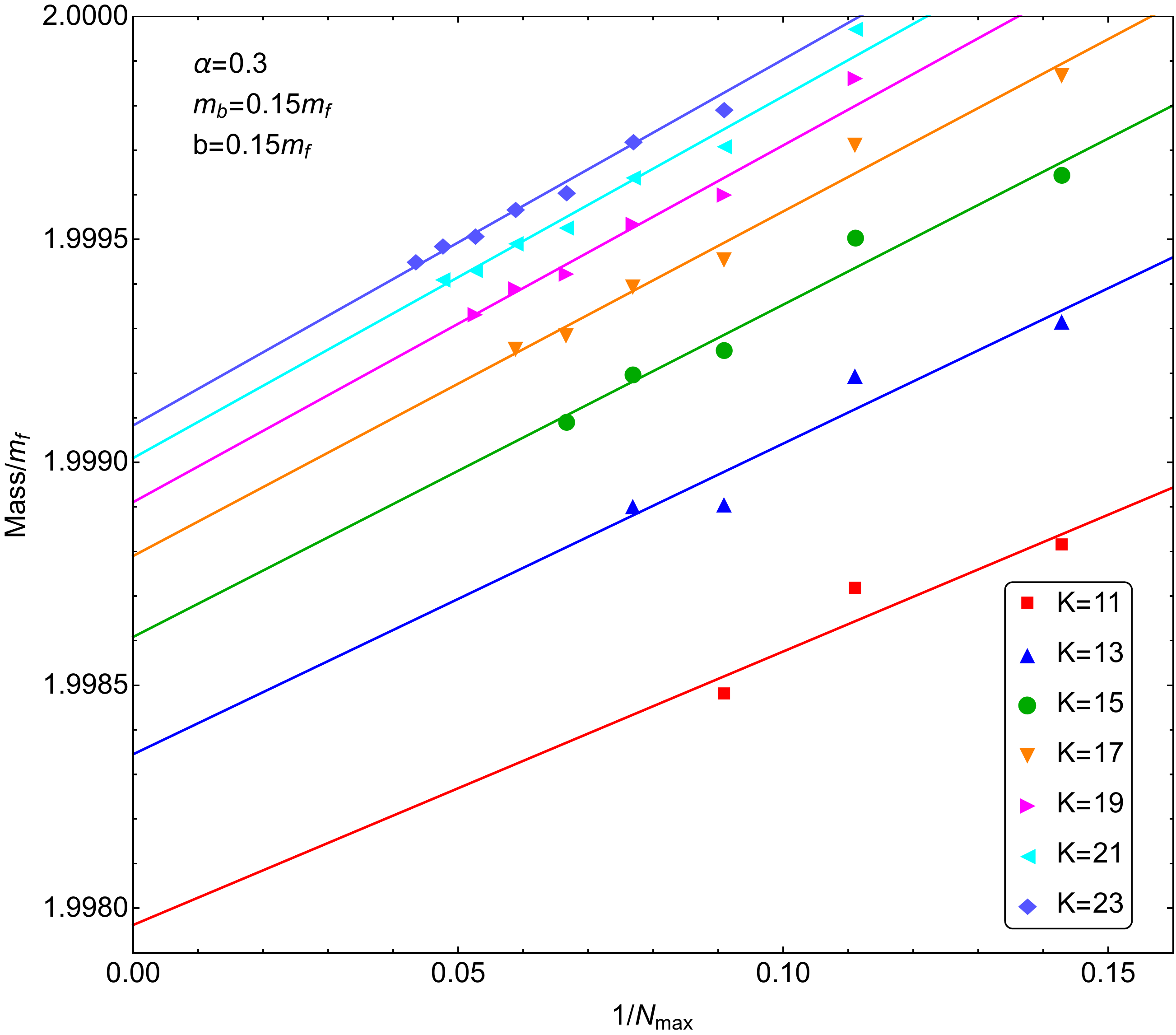}
\caption{(Color online) Mass spectrum of the Yukawa model for an $f \bar f$ system interacting
by a scalar boson exchange as a function of total magnetic projection $M_j$ 
at fixed $N_{\rm max}=K$=19 (left panel) and, for the singlet ground state, as a function of 
$1/N_{\rm max}$  at various values of $K$ (right panel). Masses are quoted in units of the fermion mass
$m_f$. Degenerate states are indicated with the ``2'' to indicate two states nearly coincide.
}
\label{Fig_Yukawa}
\end{figure}

We follow the scheme we employed in our application of BLFQ 
to positronium~\cite{Wiecki:2014ola} to solve the Yukawa model 
with a scalar boson exchange~\cite{Qian:2017aaa}.
To be specific, we solve for the mass eigenstates (in units of the fermion mass $m_f$)
as a function of two basis space
regulators, $N_{\rm max}$ and $K$ in a Fock space consisting of a fermion $f$ and an
antifermion $\bar f$ exchanging a scalar boson of mass $m_b$=0.15$m_f$
with coupling $\alpha$= 0.3 and HO length parameter $b=\sqrt{(M\Omega)}$=0.16$m_f$.
Our $f \bar f$ effective interaction follows the one employed in 
Ref.~\cite{Wiecki:2014ola} except that a scalar boson replaces the vector exchange of QED.
We neglect fermion self-energy and annihilation processes.
Our mass eigenstates are presented in units of the fermion mass in Fig. \ref{Fig_Yukawa}
for $N_{\rm max}=K=19$ (left panel) and, for the singlet ground state, as a function of 
$1/N_{\rm max}$  at various values of $K$. The spectra in the left panel is a representative case at
fixed $N_{\rm max}$ and $K$. It shows degeneracy with respect to $M_j$ whose multiplicity implies the total angular momentum for each mass according to $J=\max (|M_j|)$. Making a simple extrapolation in $1/K$ of the extrapolants obtained in the right panel with respect to $1/N_{\rm max}$, produces an estimated ground state 
mass of 1.99933 $m_f$ in the continuum limit which is close to the non-relativistic result (solution of the
Schroedinger equation with the Yukawa potential) of 1.99954 $m_f$.
Current work focuses on extending the results in the right hand panel, 
obtaining corresponding extrapolations for excited states 
and quantifying the uncertainty of the extrapolated masses.

%

\section{Heavy Quarkonium in BLFQ} \label{sec 5}

We begin here with an effective Hamiltonian $H_\text{eff}^1$, where the superscript indicates our first approximate form, based on the holographic QCD Hamiltonian  
 \cite{Brodsky09.081601}    
\begin{linenomath*}
\begin{equation}\label{eqn:Heff}
 H_\text{eff}^1 \equiv P^+\hat P^-_\text{eff} - \bm P^2_\perp 
 = \frac{\bm k^2_\perp}{x(1-x)} + \kappa^4 x(1-x)\bm r^2_\perp 
\end{equation}
\end{linenomath*}
where, $x = p^+_q/P^+$ is the longitudinal momentum fraction of the quark, $\bm k_\perp = \bm p_{q\perp} - x \bm P_\perp$ is the 
intrinsic transverse momentum, $\bm r_\perp = \bm r_{q\perp} - \bm r_{\bar{q}\perp}$ is the transverse separation of the quarks. 
$\kappa$ is the strength of the confining potential.  

The holographic QCD Hamiltonian 
is only 2-dimensional and is defined for massless quarks without interactions. For the 
heavy quarkonium systems and other applications we   
incorporate the mass of the quarks $m_q$,
add one-gluon exchange $V_\text{g}$,
and add longitudinal dynamics to arrive at our $H_\text{eff}$ \cite{Li:2015zda},
\begin{linenomath*}
 \begin{equation}
  H_\text{eff} = \frac{\bm k^2_\perp + m_q^2}{x} + \frac{\bm k^2_\perp + m_{\bar{q}}^2}{1-x} + \kappa^4 x(1-x)\bm
r^2_\perp + V_\text{g} + V_L(x)
 \end{equation}
\end{linenomath*}
Our longitudinal confining potential $V_L$ has some similarities with other
proposals \cite{Glazek11.1933,Trawinski14.074017,Chabysheva13.143}:
\begin{linenomath*}
\begin{equation}
V_{L}(x) = -\frac{\kappa^4}{(m_q+m_{\bar{q}})^2} \partial_x \big( x (1-x) \partial_x\big). 
\end{equation}
\end{linenomath*}

Among $V_L$'s appealing features are that it gives rise to 
longitudinal basis functions which are analytic solutions of the resulting 1D wave equation,
which introduces a new quantum number $\ell$,
and resemble the known asymptotic parton distribution $\sim x^\alpha (1-x)^\beta$.
In addition, in the non-relativistic limit $m_f \gg\kappa$, our longitudinal confinement combines with transverse confinement to
form a 3D-HO potential, $V_\text{con} = \frac{m_qm_{\bar{q}}}{(m_q+m_{\bar{q}})^2} \kappa^4 \bm r^2$,
and rotational symmetry is preserved. This non-relativistic reduction also provides us with simple estimates of the model parameters. 
On the other hand, in the massless limit $m_f \ll \kappa$, the longitudinal mode stays in the ground state and the longitudinal wavefunction is a constant, thereby restoring
the massless model of Brodsky and de T\'eramond \cite{Brodsky15.1}.

\begin{figure}
 \centering 
\includegraphics[width=.49\textwidth]{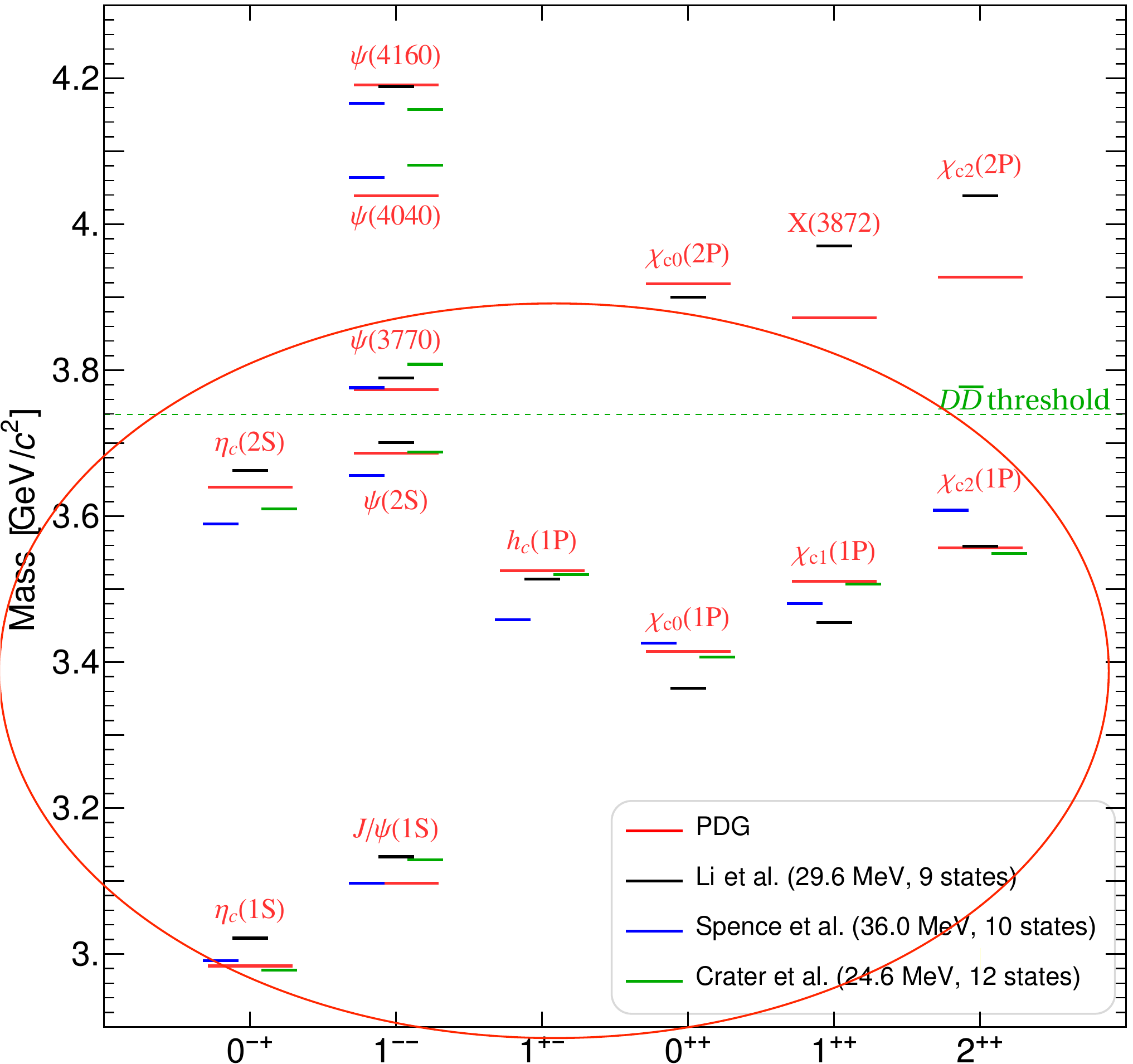}
\includegraphics[width=.49\textwidth]{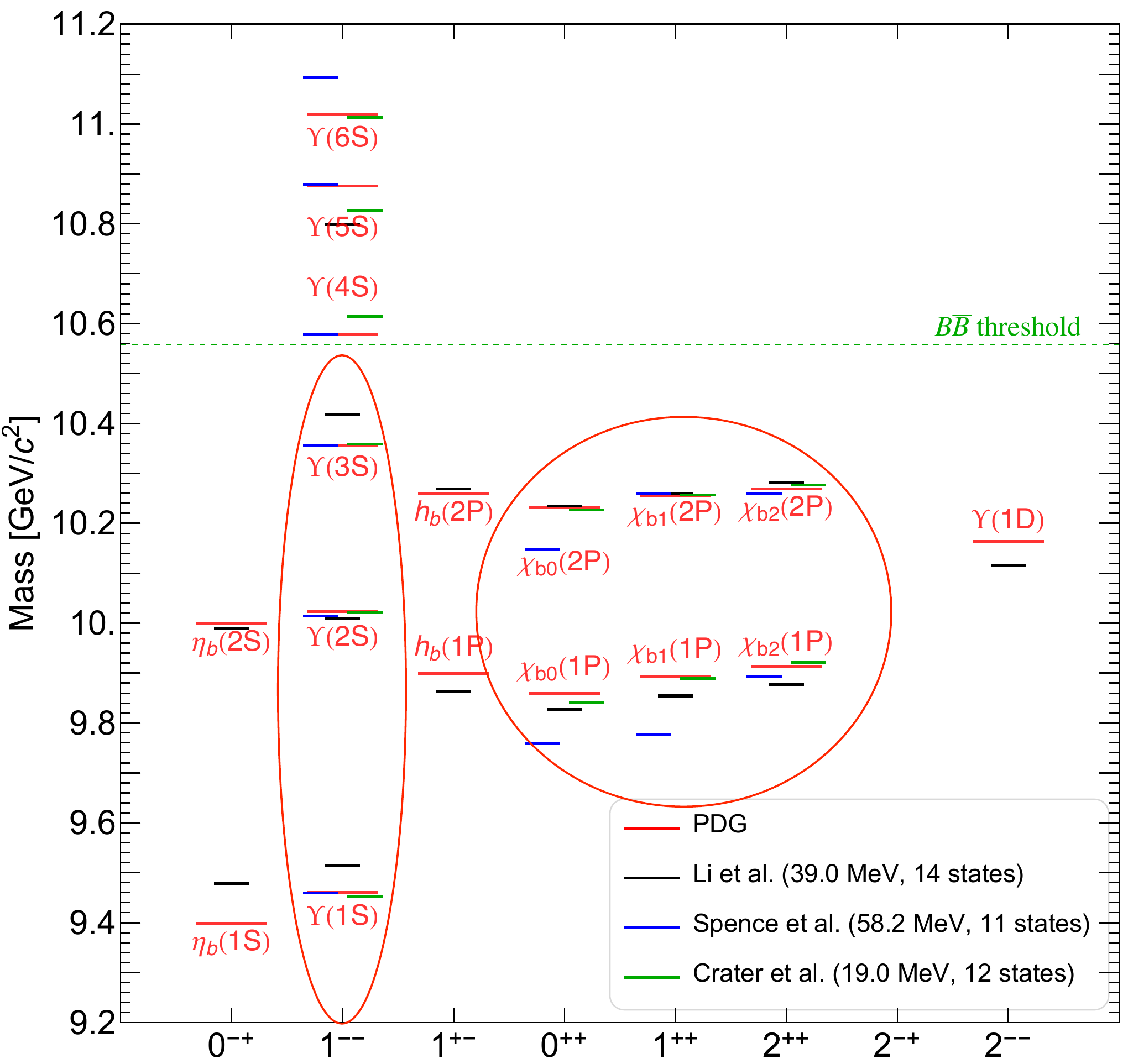}
\caption{(Color online) \textit{Left(Right)}: charmonium (bottomonium) spectrum in GeV$/c^2$;  
known states are labeled by their PDG symbols \cite{pdg.2014} while others are labeled by non-relativistic symbols. Different theoretical approaches (Refs. \cite{Li:2015zda} for``Li, et al.'', \cite{Spence2017} for ``Spence, et al.'' and \cite{Crater:2002fq} for ``Crater, et al.''), as described in the text, 
are compared with each other and with experiment (PDG). Red ovals encircle states that are quoted in all three theoretical approaches.
}
\label{fig:quark_spectrum}
\end{figure}

Our main results for heavy quarkonium are presented in more detail by Yang Li at this meeting \cite{Li_this_meeting} so we will use this opportunity to provide comparisons in 
Fig. \ref{fig:quark_spectrum} with a sample of approaches that incorporate massive quarks 
and quark-antiquark interactions. Each approach involves parameter fits but they differ widely on the choice of experimental data to incorporate in the fit and on the results presented as predictions.
The details the BLFQ approach highlighted here are presented in Ref. \cite{Li:2015zda} with more recent 
improvements presented in this meeting \cite{Li_this_meeting}. A variational approach leading to a wave
equation resembling a Bethe-Salpeter equation provides the results quoted from Ref. \cite{Spence2017}.
In this approach, confinement arises from an ansatz for the non-perturbative scale and a variational treatment of the multi-gluon dynamics that provides the shape and magnitude of the confining interaction. The two-particle Dirac equation with one gluon-exchange and confinement has been intensively investigated for all the known mesons and has led to the results shown from Ref. \cite{Crater:2002fq} that exhibit the lowest rms deviation
between theory and experiment.

We have highlighted the states that have been produced with all the cited methods by encircling them in 
Fig. \ref{fig:quark_spectrum}. Some of these states are also involved in the respective fitting procedures.
The comparisons outside the ovals provide an impression of similarities and differences in the predicted
mass eigenstates where available.  

\begin{figure}
\begin{tabular}{cc}
\subfloat[$J/\psi: 1^3S_1 (1^{--})$]{\includegraphics[scale=0.34]{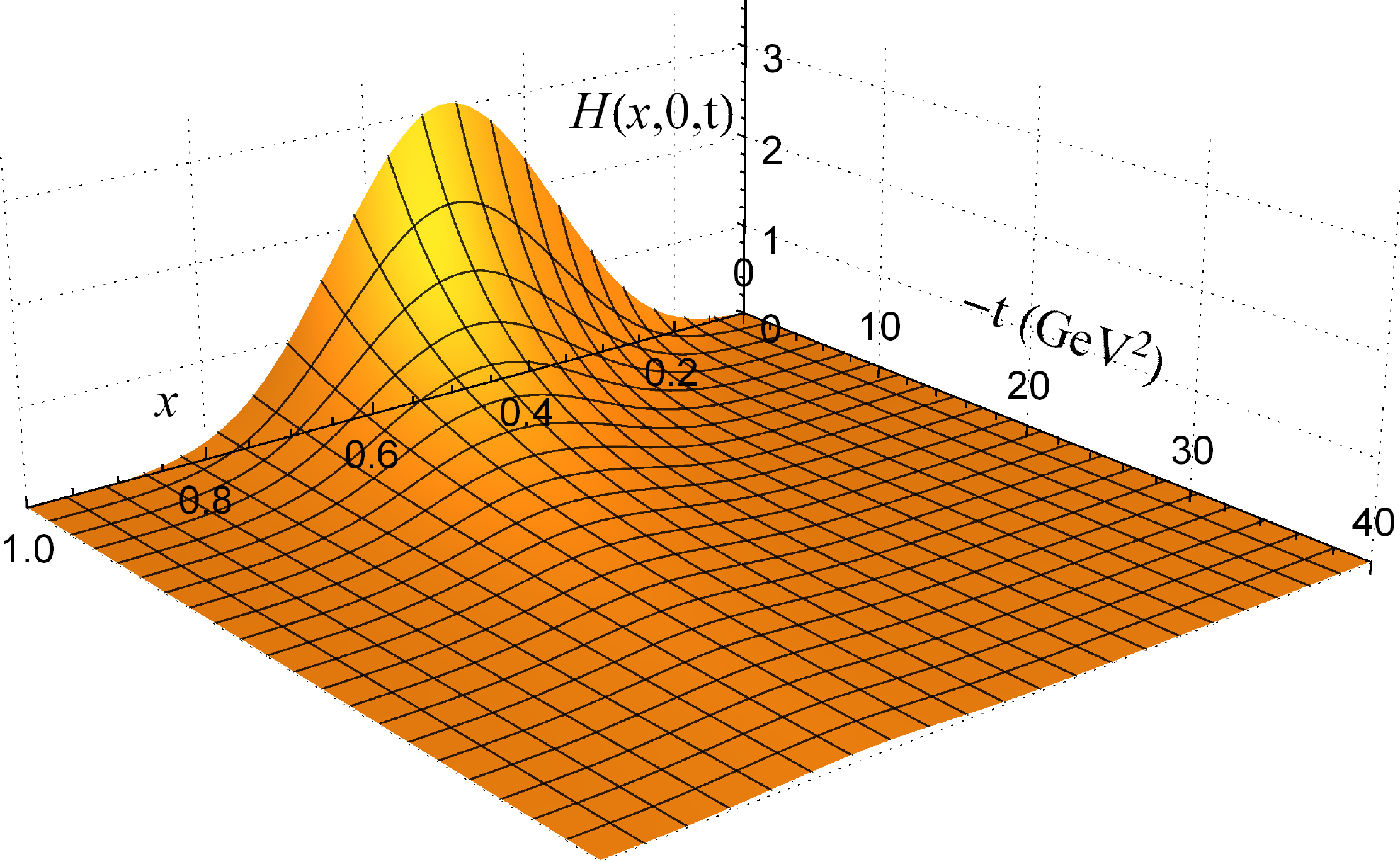}}\\
\subfloat[$J/\psi: 1^3S_1 (1^{--})$]{\includegraphics[scale=0.34]{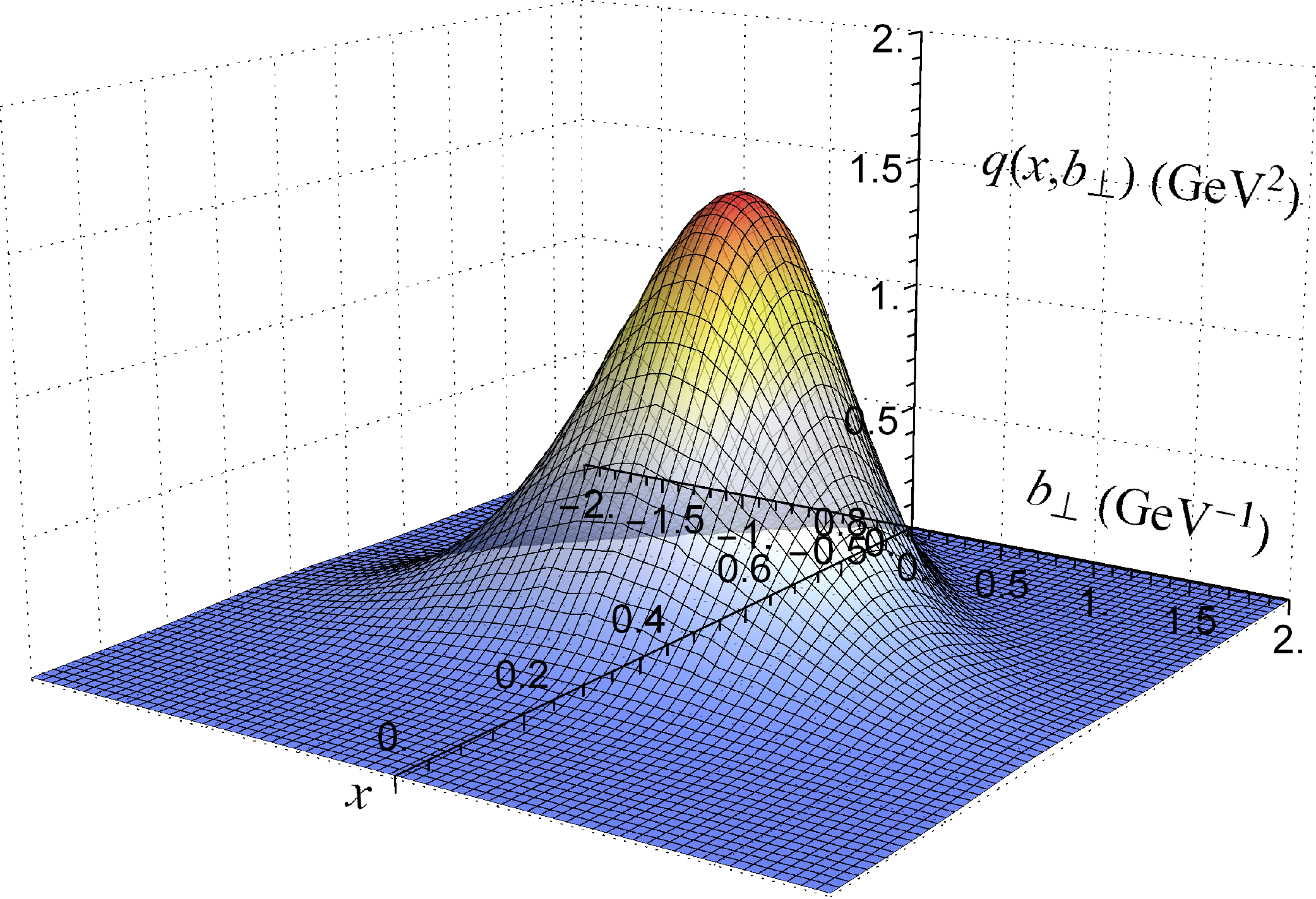}} \\
\end{tabular}
\begin{tabular}{cc}
\subfloat[$\eta^\prime_c: 2^1S_0 (0^{-+})$]{\includegraphics[scale=0.34]{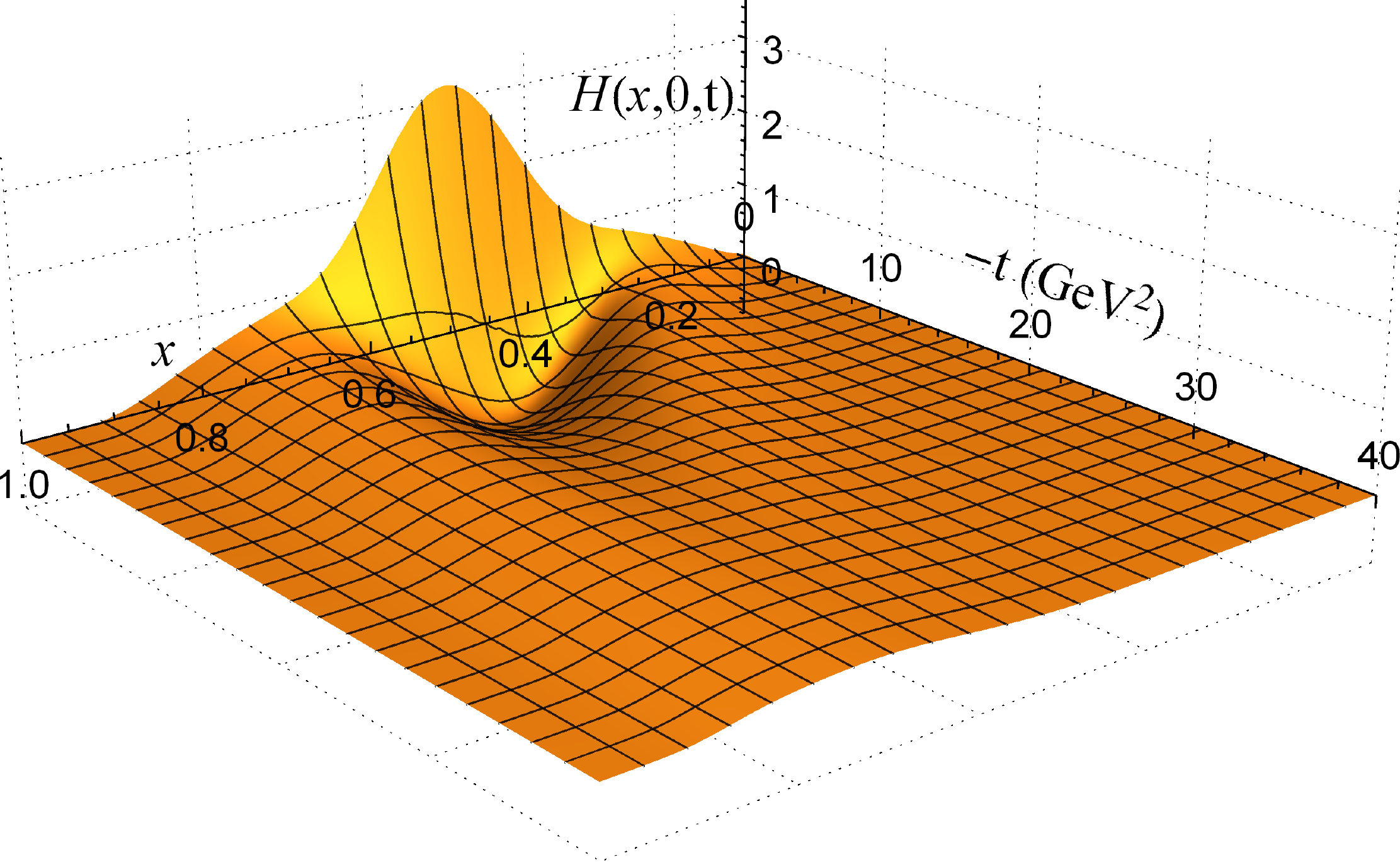}}\\
\subfloat[$\eta^\prime_c: 2^1S_0 (0^{-+})$]{\includegraphics[scale=0.34]{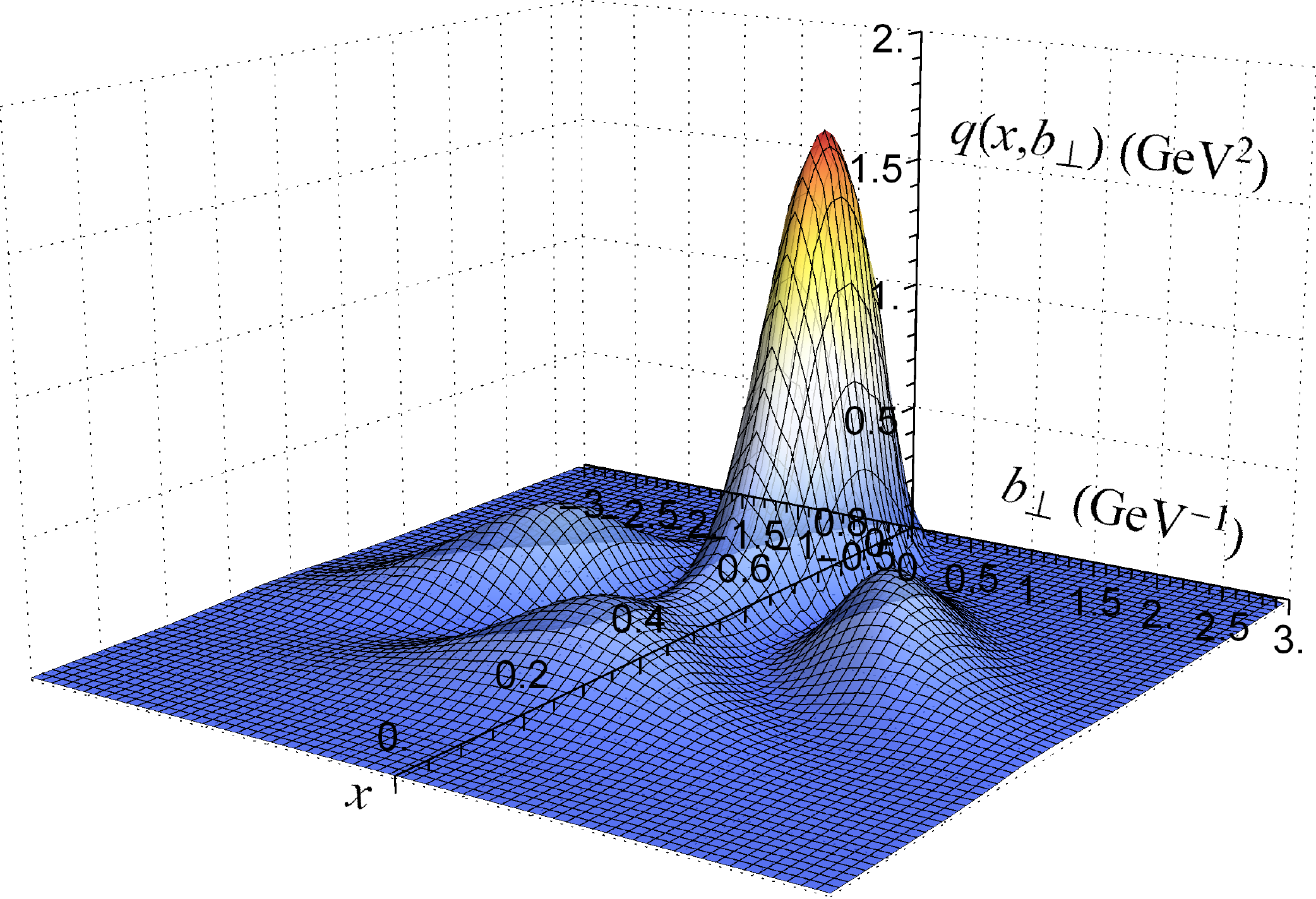}} \\
\end{tabular}
\caption{top plots: 3D plot of helicity non-flip GPDs  $H(x, \xi=0, t=-\Delta_\perp^2)$ [Eq.~\ref{eq:gpds_intrinsic}] and bottom plots: impact-parameter dependent GPDs $q(x, b_\perp)$ [Eq.~\ref{eq:gpds_space}] for the two bound states of charmonium ($c\bar{c}$) with $N_{\text{max}}=8$, $L_{\text{max}}=8$, $m_J=m_J' = 0$, coupling constant $\alpha=0.3595$, the confining strength $\kappa = 0.963$ GeV, the 2D-HO basis scale $b =\kappa $, the quark (charmonium) mass $m_q =1.492$ GeV,  and gluon mass $\mu_g = 0.02$ GeV. Note states are identified with their non-relativistic quantum numbers (relativistic quantum numbers) $n{}^{2S+1}\!L_J\, (J^{PC})$, where $n$ is the radial quantum number, the relation between $N$, the principal quantum number, and $n$, is $N=n+L$, $L$ is the total orbital angular momentum, $S$ is the total intrinsic spin, $J$ is the total angular momentum, $P$ is the parity and $C$ is the charge conjugation.}
\label{fig:gpd_b}
\end{figure}

Before closing this section, we present an example of the range of observables accessible
with our LFWFs developed in Ref. \cite{Li:2015zda}.
For brevity, we present only the helicity-non-flip 
GPD  $H(x, \xi=0, t=-{\vec \Delta}_\perp^2)$ for a selection 
of two bound states of charmonium in Fig. \ref{fig:gpd_b}. 
The helicity-non-flip GPDs can be written, for the case where the photon couples only to the quark, 
as overlap integrals between LFWFs 
\cite{Brodsky:2000xy, Frederico:2009fk, Diehl:2003ny,Brodsky:2007hb,Adhikari:2016idg}

\begin{equation}
H(x,\xi=0, t=-{\vec \Delta}_\perp^2) =  \sum_{\lambda_q,\lambda_{\bar q}}
\frac{1}{16 \pi^3 x(1-x)} \int d^2 \vec k_\perp \, \psi^*(\vec k'_\perp, x, \lambda_q,\lambda_{\bar q}) \psi(\vec k_\perp, x, \lambda_q,\lambda_{\bar q}). 
\label{eq:gpds_intrinsic}
\end{equation}

Here,  
$\bf k_\perp$ and $\bf k'_\perp $ are the respective relative transverse momenta of the quark 
before and after being struck by the virtual photon, 
$\Delta$ is the momentum transfer (we choose the Drell-Yan frame 
$\Delta^+ =0$, 
$t \equiv \Delta^2 = -{\vec \Delta}_\perp^2$, 
$\xi$ is the skewness parameter and in the Drell-Yan frame, 
$\xi=0$, and $\lambda_e (\lambda_{\bar {e}}) $ is the spin of the quark (antiquark).
This is a demonstration case illustrating hadronic GPDs in BLFQ 
similar to a positronium application~\cite{Vary:2016emi,Adhikari:2016idg} .
 
Now, referring to Ref.\cite{mb:GPD}, the impact-parameter dependent GPDs are defined as the Fourier transform of the GPDs with respect to the momentum transfer ${\vec \Delta}_\perp$
\begin{equation}  
q(x, {\vec b}_\perp) =
\int \frac{d^2{\vec \Delta}_\perp}{(2\pi)^2}
e^{-i {\vec \Delta}_\perp \cdot {\vec b}_\perp }
H(x,0,-\vec{\Delta}_\perp^2).
\label{eq:gpds_space} 
\end{equation}
Here,  the impact parameter ${\vec b}_\perp$ corresponds to the displacement of the  quark $(q)$ from the transverse center of momentum  of the entire system $(q\bar{q})$.  We present the the impact-parameter dependent GPDs for the same two states of charmonium in the lower panels of Fig. \ref{fig:gpd_b} to provide a visual impression of these coordinate space distributions. Note, especially, the appearance of secondary peaks in the $\eta^\prime_c$.

\section{Vector Meson Production} \label{sec 6}
\begin{figure}
 \centering 
\includegraphics[width=0.35\textwidth]{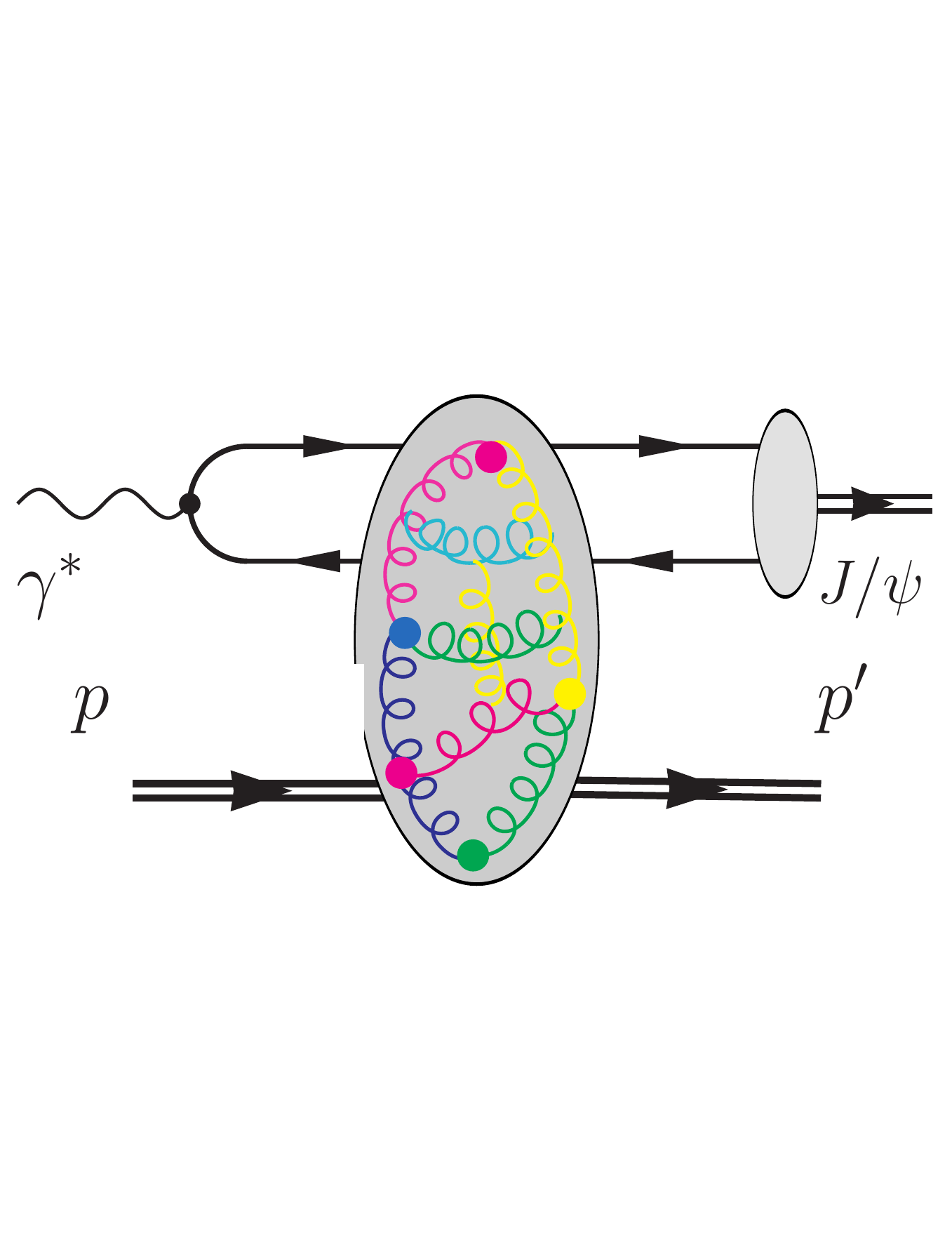}
\includegraphics[width=0.60\textwidth]{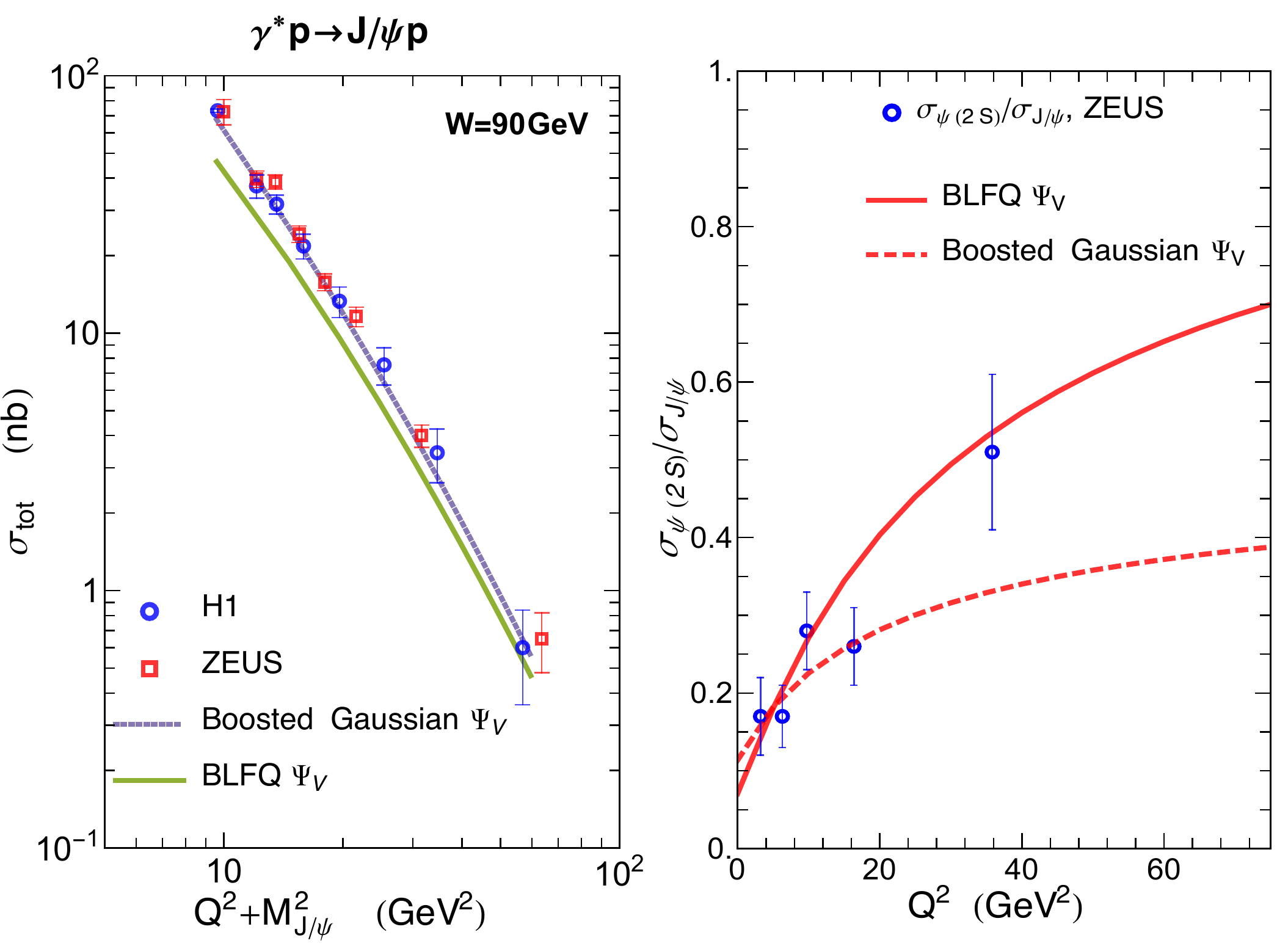}
\caption{(Color online)  
\textit{Left}: Schematic of vector meson production in the dipole model 
that begins with virtual photon dissociation to a $q \bar q$ pair, interaction with the gluon field 
of a hadron with 4-momentum $p$ and hadronization of the vector meson in the final state.
\textit{Middle}: Total $J/\Psi$ cross section for different values of $Q^2$ and $W$
\cite{Chekanov:2004mw,Aktas:2005xu}. 
\textit{Right}: Cross-section ratio $\sigma_{\Psi(2s)}/\sigma_{J/\Psi}$ for different $Q^2$.
Error bars indicate the statistical uncertainties only \cite{Abramowicz:2016ext}. }
\label{dipole_model}
\end{figure}

With our BLFQ LFWFs discussed above, we compute the production of
charmonium states in diffractive deep inelastic scattering and ultra-peripheral heavy ion collisions within the dipole picture \cite{Chen:2016dlk}. The assumed process is depicted schematically in Fig. \ref{dipole_model}
where a virtual photon dissociates into a massive $q \bar q$ pair that subsequently interacts with the 
gluon field of the hadronic system producing a vector meson in the final state. 

In the dipole model, the amplitude for  exclusive heavy quarkonium production  in DIS can be calculated as \cite{Kowalski:2006hc}
\begin{eqnarray}
  \mathcal{A}^{\gamma^* p\rightarrow Ep}_{T,L}(x,Q,\Delta) = \mathrm{i}\,\int\!d^2\bm{r}\int_0^1\!\frac{d{z}}{4\pi}\int\! d^2\bm{b}
\;(\Psi_{E}^{*}\Psi)_{T,L} (r,z,Q) \; 
  \mathrm{e}^{-\mathrm{i}[\bm{b}-(1-z)\bm{r}]\cdot\bm{\Delta}}
  \;\frac {d \sigma_{q\bar q}}{d^2 \bm b} (x,r) \; ,
  \label{eq:newampvecm}
\end{eqnarray}
where $T$ and $L$ denote the transverse and longitudinal polarization of the virtual photon (with virtuality $Q^2$) and the produced 
quarkonium, and $t= - \bm{\Delta}^2$ denotes the momentum transfer. On the right-hand side, $\bm{r}$ is the transverse size of the color
dipole, $z$ is the LF longitudinal momentum fraction of the quark, $\bm{b}$ is the impact parameter of the dipole relative to the proton and
$x$ is the Bjorken variable. $\Psi$ and $\Psi_{E}^{*}$ are LFWFs of the virtual photon, treated perturbatively,  and the exclusively produced quarkonium respectively.
The cross section is written in terms of the amplitude as 
\begin{eqnarray}
\frac{d \sigma^{\gamma^* p\rightarrow Ep}_{T,L}}{dt} = \frac{1}{16 \pi} \vert \mathcal{A}^{\gamma^* p\rightarrow
Ep}_{T,L}(x,Q,\Delta)  \vert^2 \; .
\end{eqnarray}

For the results presented in Fig. \ref{dipole_model}, 
we employ the parametrization in the impact-parameter-dependent 
saturation model provided in Ref. \cite{Kowalski:2006hc}, which is referred to as ``b-Sat I'' 
in Ref. \cite{Chen:2016dlk}. 
The calculations using the boosted Gaussian LFWF can also be found in Ref.~\cite{Kowalski:2006hc}. The results with the BLFQ LFWF (solid curves) and the phenomenological boosted Gaussian LFWF (dashed curves) both provide
reasonable descriptions of the HERA experimental data.

We obtain additional charmonium production cross sections and find reasonable agreement with
experimental data at HERA, RHIC and LHC. We observe that the cross-section ratio $\sigma_{\Psi(2s)}/\sigma_{J/\Psi}$ reveals significant independence of the gluon distribution parameters while showing
sensitivity to the charmonium LFWFs as seen in the right panel of Fig. \ref{dipole_model}. 
Additional results are found in Ref.~\cite{Chen:2016dlk}.

\begin{figure}
 \centering 
\includegraphics[width=0.75\textwidth]{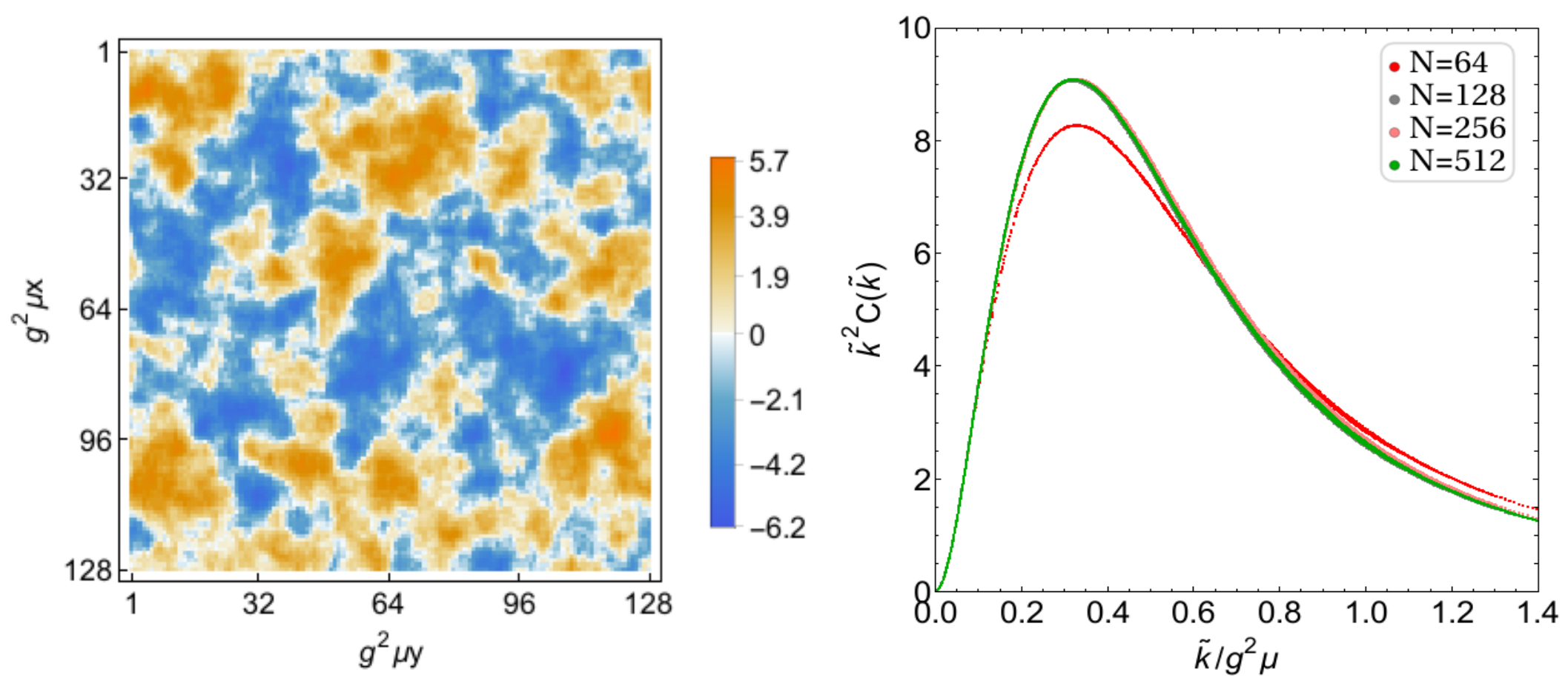}
\caption{(Color online)  
\textit{Left}: An $x-y$ slice of the amplitude of one color component of the Color Glass Condensate 
in covariant gauge with lattice size $g^2 \mu$ = 1.0 GeV, $g^2 \mu L$ = 100 and
$N$ =128 grid points on an edge.
\textit{Right}: Fourier transform of the Wilson correlator $C(k)$ times $k^2$. Each curve represents an
average over 500 color configurations with lattice spacing $g^2\mu a$ = 0.5, lattice size $L=Na$.     
}
\label{CGC_results}
\end{figure}

One of the most important uses of vector meson production is to probe the non-perturbative low-x region
of hadronic systems.  To this end, we investigate the time-dependent evolution of the $q \bar q$ system within
the Color Glass Condensate (CGC) model~\cite{McLerran:1993ni} of hadrons at low-x.  
As an initial step, we obtained the gluon field by solving the Yang-Mills Equation of stochastic color sources on a 2D lattice.  The left panel of Fig. \ref{CGC_results} shows one color component of the gluon field generated by a randomly sampled color source distribution.
To test whether the stochastic solutions represent the CGC, we examine the Wilson line correlator
$C(x_\perp - y_\perp) = \langle Tr U^\dagger(x_\perp)U(y_\perp) \rangle$ whose Fourier transform (times $k^2$)
is plotted in the right hand panel of Fig. \ref{CGC_results}.  From this we extract the saturation scale 
defined as the maximum of the curves in the right hand panel of Fig. \ref{CGC_results} as $Q^2/g^2\mu=0.32$ and observe 
good convergence with respect to the number of grid points $N$ as well as good agreement with the established CGC saturation scale~\cite{McLerran:1993ni,Lappi:2007ku}.

\section{Electron Motion in External Field with tBLFQ} \label{sec 7}

We adopt the tBLFQ framework \cite{Zhao:2013jia,Zhao:2013cma} to investigate the effects of electromagnetic (EM) fields generated by ultra-relativistic heavy ions on charged particles \cite{Chen:2016aaa}. This study is motivated by the fact that strong EM fields are generated during heavy ion collisions \cite{Tuchin:2013ie}, and a quantitative study of their effects on charged particles is essential for extracting  properties of the Quark Gluon Plasma (QGP) as well as the properties of the gluon distributions in nuclei as discussed above. Here, we investigate the real-time evolution of a quantized electron field under the influence of a strong external time-dependent EM field of a relativistic heavy ion.

\begin{figure}[ht]
\centering
\begin{minipage}[b]{0.485\linewidth}
  \includegraphics[width=0.98\textwidth,clip]{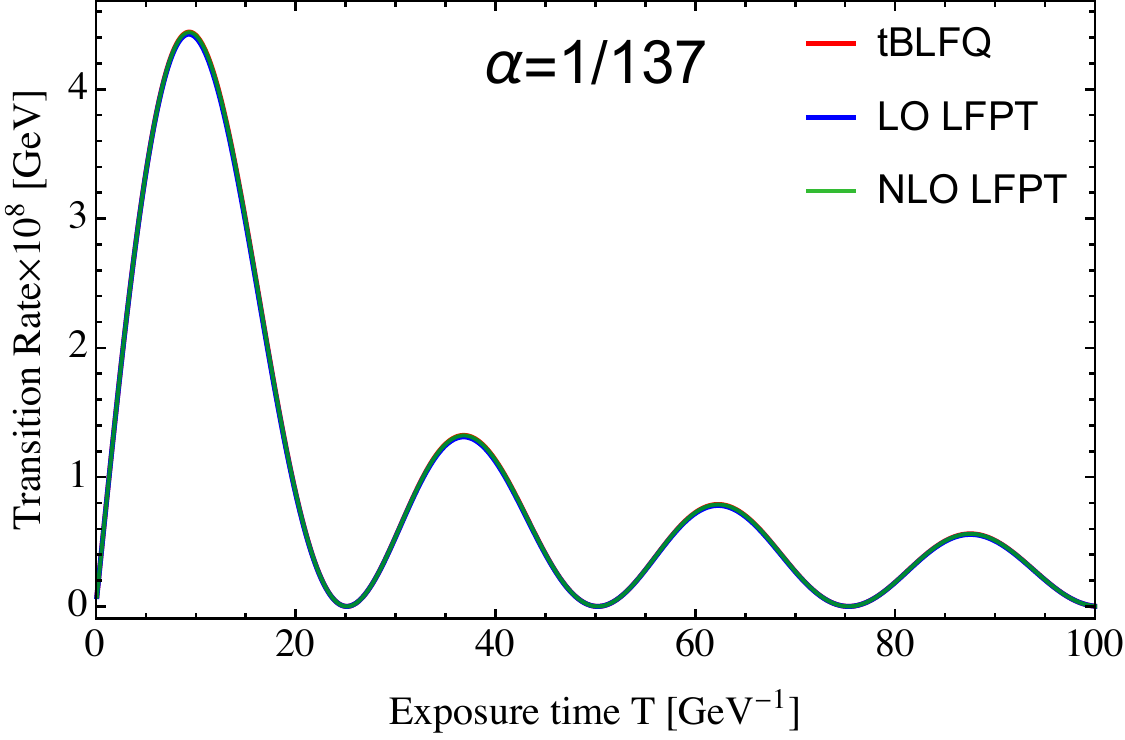}\\[2mm]
\end{minipage}
\quad
\begin{minipage}[b]{0.485\linewidth}
  \includegraphics[width=0.98\textwidth,clip]{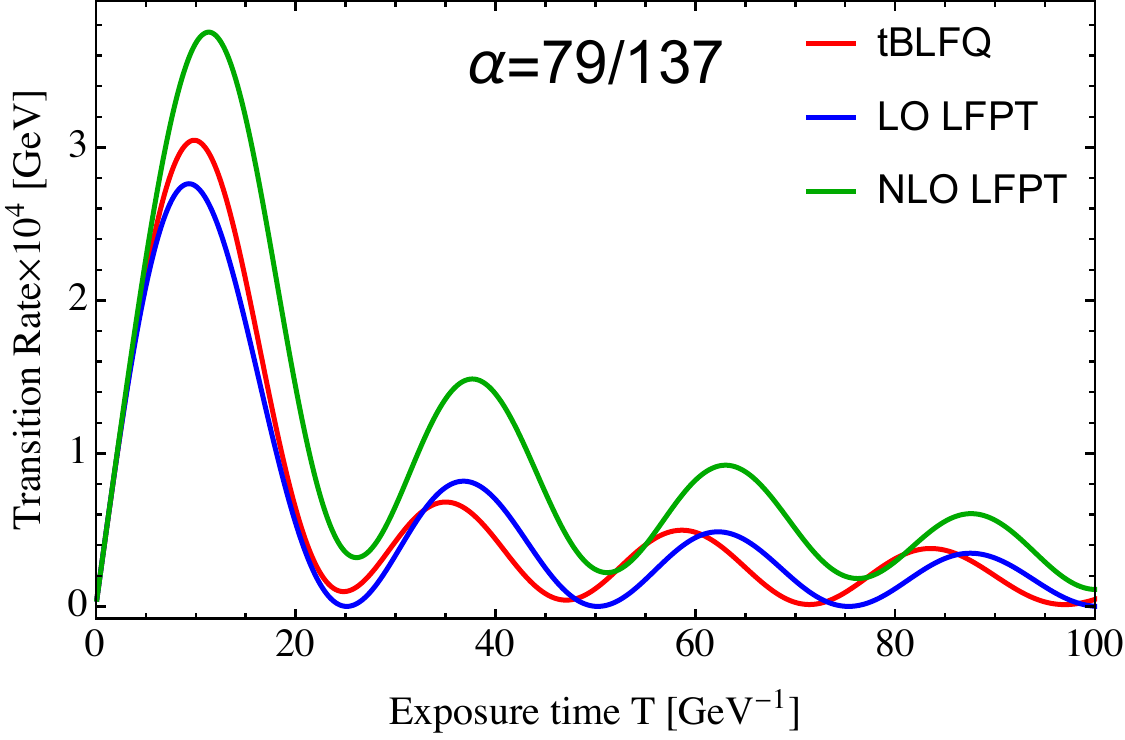}\\[2mm]
\end{minipage}
\caption{\label{fig:tBLFQ}  (Color online) Transition rate of an electron between a specific initial and a specific final state induced by the potential which is generated by nuclei with atomic number $Z=1$ (left panel) and $Z=79$ (right panel) with $\alpha \equiv Z \alpha_\text{em}$ with $\alpha_\text{em} = 1/137$. The nucleus is moving along positive $z$-axis with $y=5.3$. The initial and final states are kinetic energy eigenstates in BLFQ with energies $P^-_{\beta,i}=0.455$~GeV and $P^-_{\beta,f}=0.955$~GeV. The calculation is performed using $N_\text{max}=32$, $K=32$, $L=10$~GeV$^{-1}$ and $b=1000 m_e$,.}
\end{figure}

When the coupling between the electron and the external field is small such as the field generated by an ultra-relativistic proton with coupling $\alpha=\alpha_\text{em} \approx 1/137$, the transition rate between two kinetic energy eigenstates calculated within tBLFQ approach shows agreement with LFPT. On the other hand, for EM fields generated by an ultra-relativistic gold nucleus, the coupling between an electron and the fields is $\alpha=Z_\text{Au}\alpha_\text{em} \approx 79/137$. The transition rate between the same two kinetic energy eigenstates calculated within tBLFQ approach deviates from LFPT calculations (both leading-order (LO) and next-to-leading order (NLO)), and the differences among these calculations is a clear indication that higher order effects are significant. See Fig. \ref{fig:tBLFQ} for an illustration\footnote{The kinetic energy eigenstates are obtained through diagonalization of kinetic energy matrix in the BLFQ basis. We then evolve the electron amplitude in light-front time using the external field in the interaction picture. Fig. \ref{fig:tBLFQ} shows transition rates between two states with energy $0.0648$~GeV and $1.886$ ~GeV.}. 

\begin{figure}[ht]
\centering
\begin{minipage}[b]{0.485\linewidth}
  \includegraphics[width=0.98\textwidth,clip]{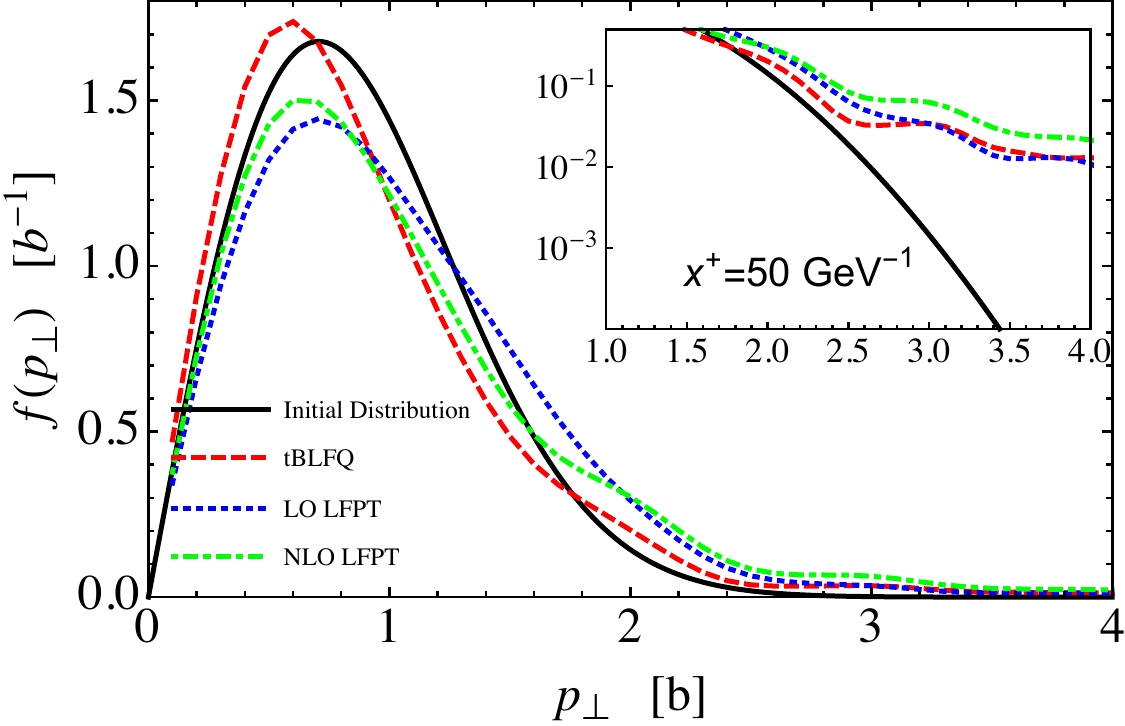}\\[2mm]
\end{minipage}
\quad
\begin{minipage}[b]{0.475\linewidth}
  \includegraphics[width=0.98\textwidth,clip]{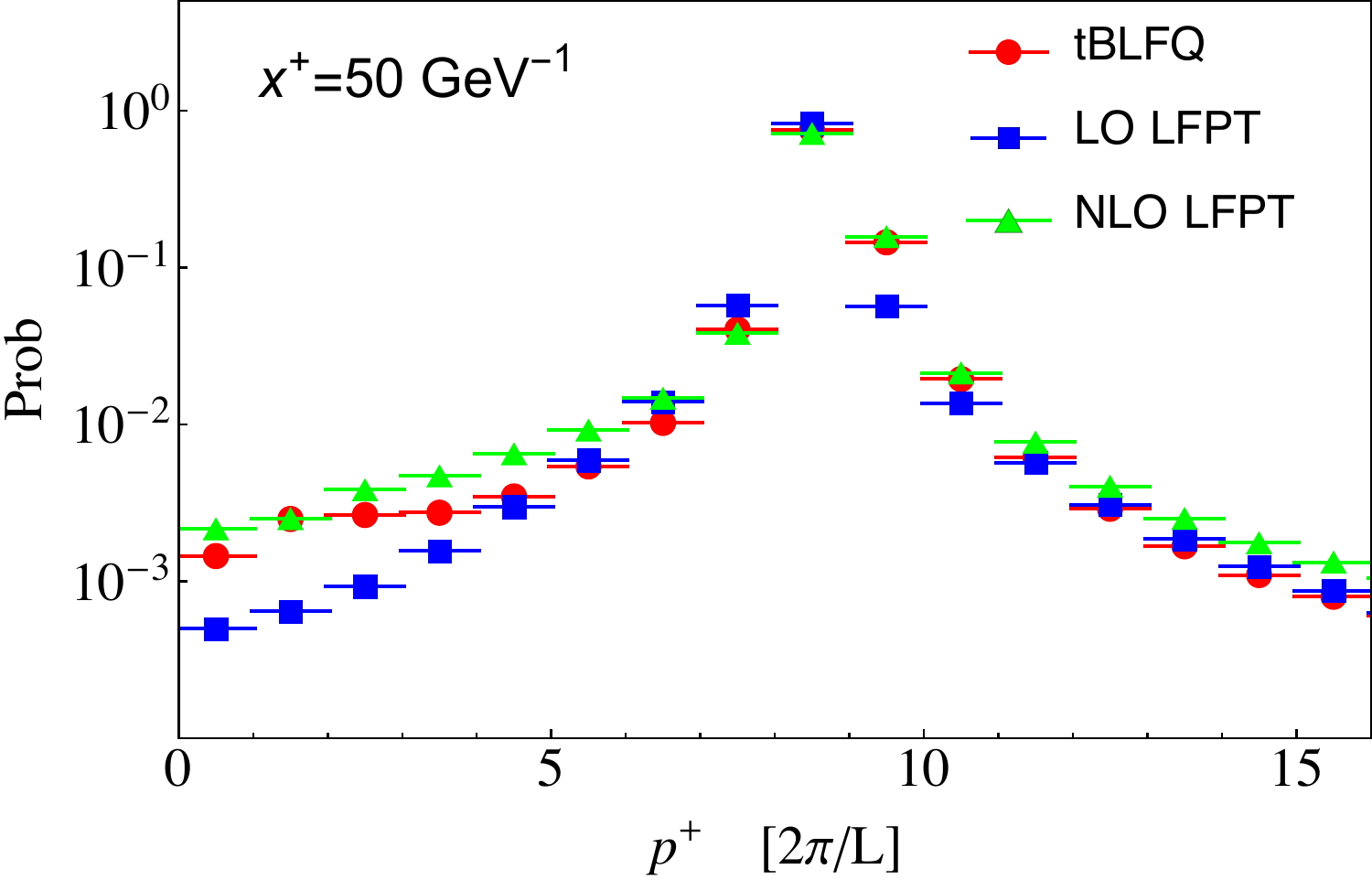}\\[2mm]
\end{minipage}
\caption{\label{fig:momentum}  (Color online) Comparisons between LFPT and tBLFQ of the
transverse (left) and longitudnal (right) momentum distribution of an electron after evolving in the EM fields
generated by a gold nucleus moving along positive $z$-axis with rapidity $y=5.3$ for $x^+ = 50$~GeV$^{-1}$. 
The initial state of the electron is the BLFQ basis state with $k=\frac{17}{2}$, $n=0$, $m=0$ and $\lambda=\frac{1}{2}$ at $x^+=0$.  
The calculation is performed using $N_\text{max}=32$, $K=32$, $L=10$~GeV$^{-1}$ and $b=1000 m_e$. 
The inset in the left panel presents a semi log plot to highlight the significant differences in the 
large transverse momentum region.
Bars in the plot of the longitudinal momentum distribution indicate the bin size for discretized plane waves in the longitudinal direction.
}
\end{figure}

We next demonstrate strong time-dependent field effects with the real-time evolution of the transverse and longitudinal momentum distributions of an electron evolving in the electromagnetic fields generated by a gold nucleus moving along positive $z$-axis with rapidity $y=5.3$. Fig.~\ref{fig:momentum} shows a snapshot of the transverse and longitudinal momentum distributions of the electron after evolving $50$~GeV$^{-1}$ (roughly $10$~fm/c) inside such strong EM fields. 
We observe that the tBLFQ simulation, which reconciles the higher order effects, predicts that the peak value of the transverse momentum will increase, accompanied by a decreased width, as compared with  the LFPT calculations predictions. See the left panel of Fig. \ref{fig:momentum} for detailed comparisons.
The probability that the electron has been excited to higher longitudinal momentum is about $35\%$ according to the tBLFQ simulation.        

Good prospects for applying the tBLFQ formalism to heavy ion collisions and electron ion collisions impel us to carry out further applications. For instance, as discussed briefly in the previous section, we will adopt the classical description of gluon fields in high energy nuclear collisions from the Color Glass Condensate (CGC) effective theory \cite{McLerran:1993ni}, and study the real-time evolution of colored objects interacting with these classical gluon fields. Within the tBLFQ framework, we can study the effects of initial gluon fields generated in the initial stage of relativistic heavy-ion collisions on heavy quarks and jets. The advantages of tBLFQ framework are distinctive: the calculation is both relativistic and at the amplitude level thereby incorporating quantum interference effects. In addition, we can naturally extend our calculation to higher Fock sectors and go beyond the Eikonal approximation.        

\section{Summary and Outlook} \label{sec 8}
It is reasonable to expect that the interplay between nuclear and hadronic physics  
to continue to flourish owing to the common challenges of strong interactions, the complexities 
of many-particle dynamics and the need to exploit the disruptive technologies of high-performance computers.  Likewise, both subfields will benefit greatly from collaborations with applied 
mathematicians and computer scientists.

The BLFQ applications summarized here represent potentially fruitful research paths.
Beyond those outlined above, we would like to mention the extension of BLFQ to other
meson sectors (mixed flavor and light mesons), to the baryons and to multi-baryon systems
which are underway.

\begin{acknowledgements}
This work was supported by the Department of Energy under Grant Nos. DE-FG02-87ER40371 
and DESC0008485 (SciDAC-3/NUCLEI).
Computational resources were provided by NERSC, 
which is supported by the Office of Science of the U.S. DOE under Contract 
No. DE-AC02-05CH11231 and by  
an INCITE award, ``Nuclear Structure and Nuclear Reactions'', from the US DOE
Office of Advanced Scientific Computing.  This research also used
resources of the Oak Ridge Leadership Computing Facility,
which is supported by the US DOE Office of Science under Contract
DE-AC05-00OR22725. X. Zhao is supported by new faculty startup funding 
by the Institute of Modern Physics, Chinese Academy of Sciences.

\end{acknowledgements}



\begin{thebibliography}{10}
%
%
%
%

\bibitem{BRANDOW:1967zz} 
  B.~H.~Brandow,
  Rev.\ Mod.\ Phys.\  {\bf 39}, 771 (1967).

\bibitem{Barrett:2013nh} 
  B.~R.~Barrett, P.~Navratil and J.~P.~Vary,
  Prog.\ Part.\ Nucl.\ Phys.\  {\bf 69}, 131 (2013).

\bibitem{Roth:2010bm} 
  R.~Roth, T.~Neff and H.~Feldmeier,
  Prog.\ Part.\ Nucl.\ Phys.\  {\bf 65}, 50 (2010).

\bibitem{Dirac:1949cp} 
  P.~A.~M.~Dirac,
  Rev.\ Mod.\ Phys.\  {\bf 21}, 392 (1949).

\bibitem{Brodsky98.299}
 S.J. Brodsky, H.-C. Pauli and S. Pinsky, 
Phys. Rept. \textbf{301}, 299 (1998).

\bibitem{Glazek:1993rc} 
  S.~D.~Glazek and K.~G.~Wilson,
  Phys.\ Rev.\ D {\bf 48}, 5863 (1993).

\bibitem{Wegner1994}
F.~Wegner, 
Ann. Phys., {\bf 506}: 77Ð91 (1994).

\bibitem{Bogner:2007rx} 
  S.~K.~Bogner, R.~J.~Furnstahl, P.~Maris, R.~J.~Perry, A.~Schwenk and J.~P.~Vary,
  Nucl.\ Phys.\ A {\bf 801}, 21 (2008).


\bibitem{Coester1958}
F.~Coester, 
Nucl. Phys. {\bf 7}, 421 (1958).

\bibitem{Coester1960}
F.~Coester and H.~Kuemmel, Nucl. Phys. {\bf 17}, 477 (1960).

\bibitem{Kuemmel2003}
H.~Kuemmel, Int. Jnl. Mod. Phys. B {\bf 17}, 5311 (2003).

\bibitem{Hagen:2013nca} 
  G.~Hagen, T.~Papenbrock, M.~Hjorth-Jensen and D.~J.~Dean,
  Rept.\ Prog.\ Phys.\  {\bf 77}, no. 9, 096302 (2014).

\bibitem{Hiller:2016itl} 
  J.~R.~Hiller,
  Prog.\ Part.\ Nucl.\ Phys.\  {\bf 90}, 75 (2016).

\bibitem{Vary:2009gt} 
  J.~P.~Vary, H.~Honkanen, J.~Li, P.~Maris, S.~J.~Brodsky, A.~Harindranath, G.~F.~de Teramond and P.~Sternberg, E. G. Ng and C. Yang,
  Phys.\ Rev.\ C {\bf 81}, 035205 (2010).

\bibitem{Honkanen:2010rc} 
  H.~Honkanen, P.~Maris, J.~P.~Vary and S.~J.~Brodsky,
  Phys.\ Rev.\ Lett.\  {\bf 106}, 061603 (2011).

\bibitem{Vary:2011np} 
  J.~P.~Vary,
  Few Body Syst.\  {\bf 52}, 331 (2012).

\bibitem{Zhao:2013jia} 
  X.~Zhao, A.~Ilderton, P.~Maris and J.~P.~Vary,
  Phys.\ Lett.\ B {\bf 726}, 856 (2013).

\bibitem{Zhao:2013cma} 
  X.~Zhao, A.~Ilderton, P.~Maris and J.~P.~Vary,
  Phys.\ Rev.\ D {\bf 88}, 065014 (2013).

\bibitem{Bakker2013.165} 
  Bakker B.~L.~G., {\it et al.}:
 \newblock Nucl.\ Phys.\ Proc.\ Suppl.\  {\bf 251-252}, 165 (2014).

\bibitem{Brodsky:2000xy}
S.J. Brodsky, M. Diehl, D.S. Hwang,
Nucl. Phys. B \textbf{596}:99--124 (2001).

\bibitem{Afanasiev:2009hy} 
  S.~Afanasiev {\it et al.} [PHENIX Collaboration],
  Phys.\ Lett.\ B {\bf 679}, 321 (2009).

 \bibitem{Abbas:2013oua} 
  E.~Abbas {\it et al.} [ALICE Collaboration],
  Eur.\ Phys.\ J.\ C {\bf 73}, no. 11, 2617 (2013).

\bibitem{Navratil:2000ww} 
  P.~Navratil, J.~P.~Vary and B.~R.~Barrett,
  Phys.\ Rev.\ Lett.\  {\bf 84}, 5728 (2000).

\bibitem{Navratil:2000gs} 
  P.~Navratil, J.~P.~Vary and B.~R.~Barrett,
  Phys.\ Rev.\ C {\bf 62}, 054311 (2000).

\bibitem{Maris:2008ax} 
  P.~Maris, J.~P.~Vary and A.~M.~Shirokov,
  Phys.\ Rev.\ C {\bf 79}, 014308 (2009).

\bibitem{Li:2013cga}
Yang Li, P.W. Wiecki, X. Zhao, P. Maris, J.P. Vary, 
Proc. Int. Conf. Nucl. Theor. Supercomputing Era (NTSE-2013), Ames, IA, USA, May 13-17, 2013. 
Eds. A.M. Shirokov and A.I. Mazur. Pacific National University, Khabarovsk, 
Russia, 2014, p. 136. 

\bibitem{Maris:2013qma} 
  P.~Maris, P.~Wiecki, Y.~Li, X.~Zhao and J.~P.~Vary,
  Acta Phys.\ Polon.\ Supp.\  {\bf 6}, 321 (2013).

\bibitem{Brodsky09.081601}
G.F. de Teramond, S.J. Brodsky, 
Phys. Rev. Lett. \textbf{102}, 081601 (2009).

\bibitem{Brodsky15.1}
S.J. Brodsky, G.F. de Teramond, H.G. Dosch, J. Erlich,
 Phys.~Rept. \textbf{584}, 1 (2015).

\bibitem{Zhao:2014xaa} 
  X.~Zhao, H.~Honkanen, P.~Maris, J.~P.~Vary and S.~J.~Brodsky,
  Phys.\ Lett.\ B {\bf 737}, 65 (2014).

\bibitem{Wiecki:2014ola}    
P.~Wiecki, Y.~Li, X.~Zhao, P.~Maris and J.~P.~Vary,   
  Phys.\ Rev.\ D {\bf 91}, no. 10, 105009 (2015).

\bibitem{Tang:2017aaa}
S.~Tang, et al., in preparation.

\bibitem{Qian:2017aaa}
W.~Qian, et al., in preparation.

\bibitem{Li:2015zda} 
  Y.~Li, P.~Maris, X.~Zhao and J.~P.~Vary,
  Phys. Letts. B {\bf 758}, 118 (2016);

\bibitem{Chen:2016aaa} 
  G.~Chen, X.~Zhao, Y.~Li, P.~Maris, K.~Tuchin and J.~P.~Vary, 
in preparation.  

\bibitem{Zhao:2014hpa} 
  X.~Zhao,
  Few Body Syst.\  {\bf 56}, no. 6-9, 257 (2015).

\bibitem{Zhao_this_meeting}
  X.~Zhao, invited talk at this meeting.

\bibitem{Brodsky:2000ii} 
  S.~J.~Brodsky, D.~S.~Hwang, B.~Q.~Ma and I.~Schmidt,
  Nucl.\ Phys.\ B {\bf 593}, 311 (2001).

\bibitem{pdg.2014}
K.A. Olive \textit{et al.} (Particle Data Group), 
Chin. Phys. C38, 090001 (2014);
[\texttt{http://pdg.lbl.gov}].

\bibitem{Spence2017}
J.~R.~Spence and J.~P.~Vary, in preparation.

\bibitem{Crater:2002fq} 
  H.~Crater and P.~Van Alstine,
  Phys.\ Rev.\ D {\bf 70}, 034026 (2004).

\bibitem{Glazek11.1933}
S.D. G\l{}azek, 
Acta Phys. Polon. \textbf{B42}, 1933 (2011).

 \bibitem{Trawinski14.074017}
 A.P. Trawi\'nski, S.D. G\l{}azek, S.J. Brodsky, 
G.F. de T\'eramond, H.G. Dosch,
Phys. Rev. D \textbf{90}, 074017 (2014).

\bibitem{Chabysheva13.143}
S.S. Chabysheva, and J.R. Hiller,
Ann. of Phys. \textbf{337}, 143-152 (2013).

\bibitem{Li_this_meeting}
Y.~Li, paper in these proceedings.

\bibitem{Frederico:2009fk}
  T.~Frederico, E.~Pace, B.~Pasquini and G.~Salme,
  Phys.\ Rev.\ D {\bf 80}, 054021 (2009).

  \bibitem{Diehl:2003ny}
  M.~Diehl,
  Phys.\ Rept.\  {\bf 388}, 41 (2003).

\bibitem{Brodsky:2007hb} 
  S.~J.~Brodsky and G.~F.~de Teramond,
  Phys.\ Rev.\ D {\bf 77}, 056007 (2008).

\bibitem{Vary:2016emi} 
  J.~P.~Vary, L.~Adhikari, G.~Chen, Y.~Li, P.~Maris and X.~Zhao,
  Few Body Syst.\  {\bf 57}, no. 8, 695 (2016).

\bibitem{Adhikari:2016idg} 
  L.~Adhikari, Y.~Li, X.~Zhao, P.~Maris, J.~P.~Vary and A.~A.~El-Hady,
  Phys. Rev. C \textbf{93}, 055202 (2016).

 \bibitem{mb:GPD}
  M.~Burkardt,
  Phys.\ Rev.\ D {\bf 62}, 071503 (2000);
  \textit{Erratum}:\textit{ibid}.~{\bf 66}, 119903(E) (2002).

\bibitem{Chen:2016dlk} 
  G.~Chen, Y.~Li, P.~Maris, K.~Tuchin and J.~P.~Vary,
  arXiv:1610.04945 [nucl-th].
 
\bibitem{Kowalski:2006hc} 
  H.~Kowalski, L.~Motyka and G.~Watt,
  Phys.\ Rev.\ D {\bf 74}, 074016 (2006).

\bibitem{Chekanov:2004mw} 
  S.~Chekanov {\it et al.} [ZEUS Collaboration],
  Nucl.\ Phys.\ B {\bf 695}, 3 (2004).
  
\bibitem{Aktas:2005xu} 
  A.~Aktas {\it et al.} [H1 Collaboration],
  Eur.\ Phys.\ J.\ C {\bf 46}, 585 (2006).

\bibitem{Abramowicz:2016ext} 
  H.~Abramowicz {\it et al.} [ZEUS Collaboration],
  PoS DIS {\bf 2015}, 078 (2015).

\bibitem{McLerran:1993ni} 
  L.~D.~McLerran and R.~Venugopalan,
  Phys.\ Rev.\ D {\bf 49}, 2233 (1994).

\bibitem{Lappi:2007ku} 
  T.~Lappi,
ÊÊEur.\ Phys.\ J.\ C {\bf 55}, 285 (2008).
ÊÊ[arXiv:0711.3039 [hep-ph]].

\bibitem{Tuchin:2013ie} 
  K.~Tuchin,
  Adv.\ High Energy Phys.\  {\bf 2013}, 490495 (2013).
   

\end{thebibliography}
\end{document}